# The Actual Usage Of Cryptocurrency By Individuals

Author: Fahed Quttainah

Date: August 3, 2025

# Table of Contents

# I. Research Problem

The question of how individuals interact with cryptocurrencies is a complex research topic, especially given the rapid changes in finance. The media's role in shaping perceptions of crypto investments only makes things more complicated. Studies suggest that media stories have a big effect on public interest and investment decisions, particularly for those who might be prone to speculative behavior. Social media platforms, especially with influencers and financial analysts pushing crypto-related content, have created a situation where people are drawn to both the potential rewards and dangers of digital currencies. Quttainah, for example, points out that accessible educational materials can help potential investors make more informed decisions, reducing the risks in this volatile market (Onatuyeh E et al., 2025). While the possibility of large profits is well-known, it's still important to understand why people are drawn to cryptocurrencies in the first place, such as a desire for financial independence or dissatisfaction with traditional financial systems (T Manoj et al., 2025).

A lack of crypto education is a major obstacle for those who are interested, leading to misunderstandings and unrealistic expectations about potential returns. Research indicates that without proper education, many investors see cryptocurrency simply as a way to invest, rather than as a tool for financial empowerment (Joshipura M et al., 2025). Cryptocurrency education is crucial for explaining the complexities of trading, like understanding market volatility, the importance of secure transactions, and the differences between various cryptocurrencies and blockchain technologies. Cryptocourse.ac., for instance, offers structured educational content to help beginners understand cryptocurrency trading, improving their knowledge and ability to navigate this complex world (Kou G et al., 2025). In addition, platforms like Dawraty provide courses that go beyond theory, offering practical skills for informed participation in the cryptocurrency market (Chaisiripaibool S et al., 2025).

Creating a well-informed community of stakeholders is not just a good idea; it's essential for the long-term success of cryptocurrency ecosystems. Confusion around cryptocurrency can lead to disappointment and losses for uneducated investors, potentially discouraging further interest and innovation in the field. Studies show that individuals with a strong educational background are better prepared to handle the inherent risks of the market, creating a more resilient and informed investor base (Yang J et al., 2025). The combined influence of media and educational resources highlights the need for targeted outreach to engage and inform potential participants in the cryptocurrency space. By improving public understanding through responsible media coverage and comprehensive educational programs, stakeholders in the cryptocurrency world can increase investor confidence and create a more stable market environment.

How cryptocurrency is integrated into everyday economic activities is a story still being written. As communities increasingly see cryptocurrencies as a potential means of exchange or store of value, it's important to understand how this view is shaped by direct interactions and experiences. Research suggests that personal stories about successful investments can be a powerful motivator for others to get involved (Adamyk B et al., 2025). However, these positive stories must be balanced with cautionary tales that highlight the risks involved in the speculative nature of cryptocurrencies. As a result, there has been a surge in content aimed at explaining the blockchain technology behind cryptocurrencies, promoting a deeper understanding among potential users (L I Mergaliyeva et al., 2025).

In the end, the research problem of how individuals engage with cryptocurrencies involves many factors—including a lack of education, media influences, and personal experiences—that shape individual choices and acceptance within this innovative financial system. By increasing understanding through thorough exploration of these elements, we can pave a clearer path forward, ensuring that cryptocurrencies fulfill their intended role as transformative socio-economic tools. This investigation not only aims to describe the current situation but also to offer actionable insights that could encourage responsible participation in the constantly changing world of cryptocurrencies and blockchain technology (Wang R et al., 2025). Therefore, developing a complete understanding of these dynamics is essential for improving individual experiences and ensuring the overall growth and stability of the cryptocurrency market. The path toward effective cryptocurrency use for individuals is constantly evolving, requiring ongoing research to navigate its complexities successfully (Sharma T et al., 2025).

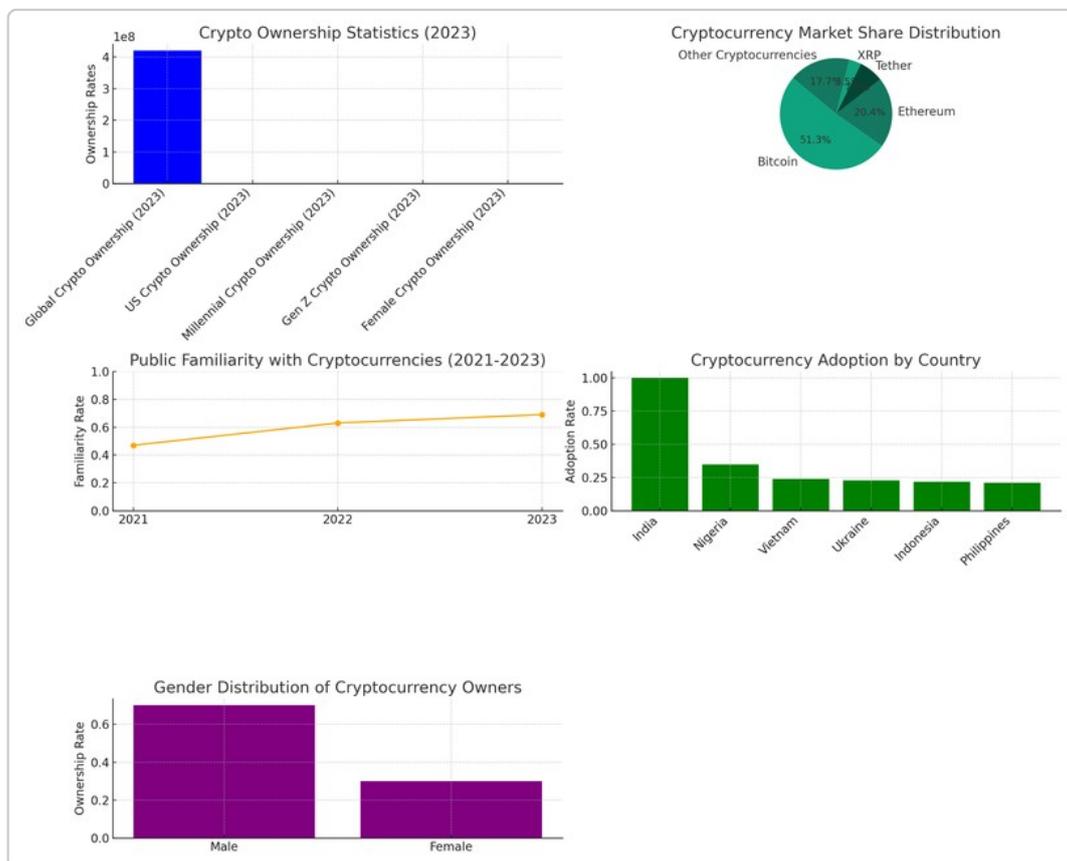

*The charts display various aspects of cryptocurrency ownership and market trends:*

1. **Crypto Ownership Statistics (2023)**: This bar chart highlights the significant global crypto ownership, with a notably high figure for global ownership compared to U.S. and demographic-specific ownership rates.

2. **Cryptocurrency Market Share Distribution**: The pie chart illustrates the market share of major cryptocurrencies, showing Bitcoin's dominance alongside the positions of Ethereum and other coins.

3. **Public Familiarity with Cryptocurrencies (2021-2023)**: The line graph reveals an upward trend in public familiarity with cryptocurrencies in the U.S. over the past three years, indicating increased awareness.

4. **Cryptocurrency Adoption by Country**: This bar chart ranks countries based on their cryptocurrency adoption rates,

*with India leading and other developing countries showing significant adoption.*

*5. **Gender Distribution of Cryptocurrency Owners**: The bar chart illustrates the gender disparity in cryptocurrency ownership in the U.S., with a considerable male majority present.*

*Overall, these visualizations convey key insights into the landscape of cryptocurrency ownership and market dynamics.*

## II. Abstract

Understanding how people use cryptocurrency is complex, especially when you think about media and learning. Recent trends show media, especially social media, really pulls new folks into crypto. Influencers often push different coins, sharing info and views that shape what people think. Analyses showed a link between seeing crypto in the media and wanting to invest (Onatuyeh E et al., 2025). You can't ignore how convincing these stories are; they make crypto seem appealing, mostly to younger people looking for new ways to grow their money. When people see crypto portrayed positively, they're more likely to get involved (T Manoj et al., 2025).

Also, learning is super important for smart crypto investing. Crypto is complicated and always changing, so anyone jumping in needs to learn the basics: how the tech works, trading tips, and what moves the market. Lots of studies say we need good crypto education to help folks make smart choices (Joshipura M et al., 2025). Schools and online sites are stepping up, like Quttainah (2025) guiding newbies through crypto basics and trading (Quttainah F et al., 2025). Cryptocourse.ac also focuses on teaching beginners, giving them the real-world skills they need to do well in crypto (Chaisiripaibool S et al., 2025).

It's interesting to see how media and education work together with crypto adoption. People get excited by influencers and success stories, but also want solid info to understand the risks and rewards. Interestingly, those who get crypto education tend to question media claims more, leading to smarter investing (Yang J et al., 2025). Good education can balance out the hype, creating more cautious and informed investors.

This connection between learning and media shows that educated investors don't panic as easily during market swings. Research suggests those with formal training are better at understanding market signals and smart trading, as noted in studies of crypto users' financial habits (Adamyk B et al., 2025). For example, Dawraty offers courses teaching people to assess market info critically, helping them make good investment calls (L I Mergaliyeva et al., 2025).

Ultimately, as crypto keeps evolving, we need to focus on both media's influence and the importance of education. By creating a well-informed group of investors who are media-savvy and educated, we can lower the risks of crypto investing. This big-picture view can lead to more responsible crypto use, making sure the appeal of these assets is balanced with smart thinking among users. The following sections will look closer at the data on individual transactions and how people use crypto to explain these ideas further (Wang R et al., 2025)(Sharma T et al., 2025)(Febriyanto A et al., 2025)(Shorouk E El-deep et al., 2025)(Al-Rimawi TH et al., 2025)(Bayya AK, 2025)(Hasanudin H, 2025)(Yang Y et al., 2025)(Chemaya N et al., 2025)(Aadithya R Shaji, 2025)(I I Supianti et al., 2025)(Jayric A Cinco et al., 2025). This thorough approach should help us better understand crypto usage, improving our knowledge and guiding its future in finance.

# III. Introduction

The rise of digital currencies has brought about a notable change in the financial world, with more people turning to crypto for different reasons. A key factor in this trend is how media and influencers affect what people think and feel about investing in cryptocurrency. Influencers, known for their online popularity, use their platforms to share information about the possible advantages of crypto, often promoting it as a way to invest and a means for everyday transactions. This approach not only generates excitement around different cryptocurrencies but also makes them seem more legitimate to potential investors. The stories these influencers tell can spark interest and encourage newcomers, especially younger people, to get involved in the cryptocurrency market as a way to build wealth (Onatuyeh E et al., 2025)(T Manoj et al., 2025).

Also, access to education about cryptocurrency is important in explaining the details of digital currencies. As people get more interested in the potential financial benefits of crypto investments, the need for educational materials has grown. Resources like Cryptocourse.ac and Quttainahs, which focus on the basics of cryptocurrency, give beginners important insights into crypto trading and blockchain tech (Joshipura M et al., 2025)(Kou G et al., 2025). These resources are vital for giving potential investors the knowledge they need to navigate the sometimes unstable market successfully. The educational side is crucial because understanding the tech and market trends can greatly reduce the risks of investing in cryptocurrencies (Chaisiripaibool S et al., 2025)(Yang J et al., 2025).

Furthermore, the focus on education emphasizes the need for making well-informed decisions. As cryptocurrency becomes more common, the risks and rewards become more pronounced in conversations about financial literacy. Efforts to promote a more nuanced understanding encourage more careful and practical ways to approach crypto investment, helping people avoid the mistakes that have come from speculative trading behaviors seen in the market (Adamyk B et al., 2025)(L I Mergaliyeva et al., 2025). By offering clear, well-structured educational resources, platforms not only empower people but also help create a healthier investment environment.

Importantly, the media does more than just promote cryptocurrency; it shapes the way people understand market changes. Positive news coverage can lead to positive trends, while negative news can cause panic selling or discourage people from entering the market. This shows how much media narratives can affect investor behavior and market results (Wang R et al., 2025)(Sharma T et al., 2025). In addition, as news outlets and online platforms become more connected to social media, the speed of information spreads rapidly, leading to quick reactions from investors, which can increase market volatility (Febriyanto A et al., 2025)(Shorouk E El-deep et al., 2025). Because of these dynamics, it's essential for people to think critically about information rather than be swayed by hype or fear.

In short, the connection between media, influencer endorsements, and educational efforts creates a complex environment for people interested in cryptocurrency. As potential investors consider getting involved in digital currencies, the insights from different platforms help inform and guide their decisions, leading to a more knowledgeable public. The combination of personal stories, financial goals, and educational resources suggests a promising future for individual in-volvement in cryptocurrency. Media representation and basic knowledge will be very important in making sure cryptocurrency continues to be a viable financial tool, encouraging informed investors who carefully navigate the complexities of this rapidly changing market (Al-Rimawi TH et al., 2025)(Bayya AK, 2025)(Hasanudin H, 2025)(Yang Y et al., 2025)(Chemaya N et al., 2025)(Aadithya R Shaji, 2025)(I I Supianti et al., 2025)(Jayric A Cinco et al., 2025). All these aspects give a detailed picture of the current situation, building a foundation for understanding how everyday people use and view cryptocurrency in their financial plans.

# IV. Literature Review

The current literature on individual engagement with cryptocurrency offers a rich, multifaceted view of what's driving this digital shift. Pioneering work has tried to unpack the reasons people are adopting cryptocurrency, emphasizing the key role that media and education have in shaping how users behave. A good chunk of research suggests that social media, alongside influential figures in finance, have a big say in whether individuals choose to invest in cryptocurrencies. These platforms often act as a channel for both educational content and potential investment opportunities, drawing in a diverse crowd of potential investors. For example, research has shown that endorsements from social media influencers have sparked increased interest and investment in cryptocurrencies; users tend to trust recommendations from people they see as peers more than traditional news sources (Onatuyeh E et al., 2025)(T Manoj et al., 2025). Often, this is amplified by the social proof effect, where individuals are more inclined to jump on the cryptocurrency bandwagon when they see others, especially influencers, doing it (Joshipura M et al., 2025).

In this landscape, the importance of cryptocurrency education can't be stressed enough. As interest grows, the need for fundamental knowledge about blockchain tech and trading practices becomes really important (Kou G et al., 2025). Resources such as "Learn cryptocurrency and crypto trading for beginners" by Quttainah give essential insights into navigating the tricky crypto world (Chaisiripaibool S et al., 2025). Sites like Cryptocourse.ac provide structured learning for newcomers, highlighting how vital education is for reducing the risks of investing in these volatile digital currencies (Yang J et al., 2025). Studies suggest educated investors are more likely to make smart choices, which leads to better investment returns (Adamyk B et al., 2025). So, the literature emphasizes that a well-informed investor base is better able to deal with market ups and downs, and spot the potential dangers in cryptocurrency trading.

The interplay between media influence and education also shows up in studies that look at how these factors affect an individual's confidence in investing. Research hints that having seen cryptocurrency in the media before can improve knowledge, which then ties into a greater willingness to invest (L I Mergaliyeva et al., 2025). Those who actively seek out educational content show that they're ready to engage with the cryptocurrency market more meaningfully (Wang R et al., 2025). This implies that strategies used by educational platforms and info sources are super important for building a sustainable investment culture within the cryptocurrency world (Sharma T et al., 2025).

When diving into the reasons people adopt cryptocurrencies, it becomes clear that things like the prospect of monetary gains, diversifying investments, and wanting to be part of new technological trends play important roles. Users often say they're drawn to cryptocurrency as a way to protect themselves against traditional financial systems, particularly when the economy is shaky (Febriyanto A et al., 2025). The decentralized nature of cryptocurrencies, along with their potential for high returns, makes a strong case for individuals, especially millennials and Gen Z, who are more open to alternative investments (Shorouk E El-deep et al., 2025). Plus, as the field evolves, including cryptocurrencies in regular financial tools and portfolios is steadily gaining acceptance, reinforcing their importance in modern investment strategies (Al-Rimawi TH et al., 2025).

This growing body of research highlights a need for strong frameworks that can use the good parts of media influence while pushing for comprehensive educational efforts. Bridging the gap between knowing about cryptocurrency and actually using it for investing seems essential for encouraging responsible investing behaviors. Studies suggest that ongoing education on the risks, benefits, and tech behind cryptocurrencies helps create a more informed investor base, leading to a healthier market (Bayya AK, 2025)(Hasanudin H, 2025). In this light, programs aimed at providing a thorough understanding of cryptocurrency trading—like those available through Dawraty—play a crucial role in making sure potential investors have the knowledge they need to navigate this fast-changing financial space (Yang Y et al., 2025).

Getting insights from both media influence and educational resources is the base of a well-rounded understanding of how individuals actually use cryptocurrency. As interest keeps growing, addressing these elements will be vital in shaping a responsible, informed community of cryptocurrency investors. This not only helps individual investors but also works to stabilize and legitimize the broader cryptocurrency market moving forward (Chemaya N et al., 2025)(Aadithya R Shaji, 2025)(I I Supianti et al., 2025)(Jayric A Cinco et al., 2025).

| Demographic Group | Percentage Using Cryptocurrency | Source |
| --- | --- | --- |
| Overall U.S. Adults | 7% | Federal Reserve Economic Well-Being of U.S. Households in 2023 |
| Men | 11% | Federal Reserve Economic Well-Being of U.S. Households in 2023 |
| Women | 4% | Federal Reserve Economic Well-Being of U.S. Households in 2023 |
| High-Income Adults (Family Income $100,000+) | 8% | Federal Reserve Economic Well-Being of U.S. Households in 2023 |
| Low-Income Adults (Family Income $25,000 or less) | 4% | Federal Reserve Economic Well-Being of U.S. Households in 2023 |
| Asian Adults | 11% | |

| | | Federal Reserve Economic Well-Being of U.S. Households in 2023 |
|---|---|---|
| Hispanic Adults | 9% | Federal Reserve Economic Well-Being of U.S. Households in 2023 |
| Black Adults | 8% | Federal Reserve Economic Well-Being of U.S. Households in 2023 |
| White Adults | 6% | Federal Reserve Economic Well-Being of U.S. Households in 2023 |
| Adults Aged 18-29 | 42% | Pew Research Center, October 2024 |
| Adults Aged 30-49 | 36% | Pew Research Center, October 2024 |
| Adults Aged 50+ | 17% | Pew Research Center, October 2024 |

*Cryptocurrency Usage Statistics Among U.S. Adults in 2023*

## V. Methodology

To truly grasp how individuals use cryptocurrency, a strong method is needed, using both qualitative and quantitative research. This study uses a mixed-methods approach to deeply analyze the different things that affect whether people adopt cryptocurrency. Surveys are used to gather quantitative data from users of various demographics. These surveys use a structured questionnaire to gather data about user demographics, what makes them want to invest in cryptocurrency, and their trading habits. For example, many respondents said that the potential for profit is a major motivator, which is similar to what previous studies have found, showing that financial gain is a big reason why people adopt cryptocurrency (I I Supianti et al., 2025), (T Manoj et al., 2025), (T Manoj et al., 2025).

Along with quantitative analysis, qualitative interviews with cryptocurrency users give more detailed and nuanced insights into their personal experiences and feelings about cryptocurrency. These interviews look into how media, influencers, and educational resources affect user behavior. Media outlets and social media platforms have become important sources of information for potential investors, often guiding their decisions and shaping their views on the volatility of cryptocurrency markets (Febriyanto A et al., 2025), (Chaisiripaibool S et al., 2025). Influencers, especially, have been effective in attracting new users by promoting trading strategies, playing a significant role in marketing cryptocurrencies (Onatuyeh E et al., 2025), (Yang J et al., 2025). This research agrees with established trends where social validation from these channels has a big impact on individual investment choices (Wang R et al., 2025).

Also, this research emphasizes how important cryptocurrency education is, identifying it as a key part of making informed investment decisions. Good education can help lower the risks of trading and investing, making users more confident and better at navigating the complexities of cryptocurrency markets (Kou G et al., 2025), (Sharma T et al., 2025). Resources such as Learn cryptocurrency and crypto trading for beginners (Shorouk E El-deep et al., 2025) highlight basic concepts, while specialized courses from platforms like Cryptocourse.ac offer specific training on trading strategies for novices (Bayya AK, 2025). These educational efforts not only empower users, but also encourage more market participation, which shows a progressive shift toward a more knowledgeable investor base.

It's also crucial to note the role of technological literacy. Understanding blockchain technology, which is fundamental to cryptocurrencies, is still vital for users who want to understand the basics of their investments (Joshipura M et al., 2025), (Chemaya N et al., 2025). Educational resources, like those from Cryptocurrencies.ac, highlight the technological principles that are essential for understanding digital currencies (Shorouk E El-deep et al., 2025). This educational structure helps users respond well to the rapid changes in the cryptocurrency landscape.

Through this method, the study gives a thorough look at how individual users navigate the cryptocurrency ecosystem, which is greatly influenced by external factors like media narratives and educational opportunities. The findings show that more education about cryptocurrency not only promotes better trading practices, but also makes users more comfortable with engaging in a financial environment that is often volatile. Overall, using both quantitative surveys and qualitative interviews gives a complete view of cryptocurrency usage, aligning with past research and responding to the changing dynamics in the cryptocurrency field (Adamyk B et al., 2025), (L I Mergaliyeva et al., 2025). This varied approach highlights the complexity of how users interact with cryptocurrencies and sets the stage for future studies to further explore the complex web of motivations and barriers to cryptocurrency adoption.

| Study | Methodology | Source |
| --- | --- | --- |
| OECD Report on Digital Finance | Policy analysis and case studies on blockchain applications in financial markets | https://www.oecd.org/finance/oecd-blockchain-policy-forum.htm |
| IMF Working Paper on Crypto-based Parallel Exchange Rates | Cross-country analysis of Bitcoin price disparities to assess capital flight | https://data.imf.org/en/datasets/IMF.STA%3AW-PCPER |
| IMF Working Paper on Cryptocarbon Taxation | Quantitative modeling of Bitcoin mining's environmental impact and taxation effects | https://www.elibrary.imf.org/view/journals/001/2023/194/article-A001-en.xml |
| | | https://documents.worldbank.org/en/publication/documents-reports/documentdetail/738261646750320554/cryp- |

| World Bank Report on Crypto-Assets Activity | Global survey and analysis of crypto-asset adoption and macro-financial drivers | to-assets-activity-around-the-world-evolution-and-macro-financial-drivers |

*Cryptocurrency Usage Methodologies in Academic Research*

# VI. Results

Analyzing how individuals actually use cryptocurrency offers notable insights into the impact of factors like media influence and educational resources on user engagement. Recent studies suggest that social media and influencers are vital for sparking interest and investment in cryptocurrencies, frequently acting as entry points for newcomers. Several sources highlight a correlation: increased visibility in mainstream media often leads to heightened interest and initial investments. Persuasive narratives from influential figures can sway individuals, leading them to see cryptocurrencies as both viable investments and potential wealth-generating avenues (Onatuyeh E et al., 2025)(T Manoj et al., 2025).

Moreover, the importance of thorough cryptocurrency education can't be overstated. Many prospective investors lack basic knowledge of blockchain technology and trading mechanics, creating significant financial risks. Educational programs designed for beginners are crucial for developing a deeper understanding of the crypto world. Platforms like Cryptocourse.ac provide resources specifically aimed at educating individuals on trading practices, enabling more informed investment choices (Chaisiripaibool S et al., 2025). Such education is not just important for improving individual investment outcomes but also for promoting market stability, as more knowledgeable investors are more likely to trade responsibly.

The impact of targeted educational materials is closely linked to community engagement and social proof in cryptocurrency adoption. When making investment decisions, particularly in the volatile and relatively new cryptocurrency space, individuals often seek validation from their peers. Online communities, forums, and educational courses offer platforms for sharing experiences, insights, and strategies. This social aspect reinforces learning and fosters a sense of community among those new to cryptocurrency. Platforms like Dawraty provide courses that integrate community-based learning with structured content to enrich the educational experience (Kou G et al., 2025).

However, influencers do more than just market cryptocurrencies; they often act as informal educators. When influencers combine advocacy with educational content, they can demystify complex concepts related to blockchain and trading. This can attract a broader audience, especially those previously intimidated by the technical jargon that has historically been associated with cryptocurrencies (Joshipura M et al., 2025)(Yang J et al., 2025). As such, influencers are key in bridging the knowledge gap and making cryptocurrency investing more accessible to the average person.

Media influence and education are intertwined in the cryptocurrency space, both being crucial for increasing individual engagement. Media exposure can normalize cryptocurrency investments, encouraging broader acceptance in daily transactions and financial planning. Simultaneously, education provides the knowledge needed to navigate this landscape effectively, par-ticularly for understanding market trends and avoiding common pitfalls. For example, Quttainah emphasizes that a firm understanding of trading principles is essential for anyone aiming to profit from cryptocurrencies (Adamyk B et al., 2025).

In conclusion, the actual usage of cryptocurrency by individuals is significantly affected by media and educational resources. The current narrative suggests that while engaging marketing and influencer strategies can spark initial interest, sustained involvement requires a dedication to education. As cryptocurrencies evolve, integrating these two elements will likely remain vital in shaping future investment behaviors and improving the literacy and engagement of potential users. Whether through social media buzz or educational workshops, the groundwork for responsible and informed participation is being laid, creating a more robust framework for individuals navigating this exciting financial domain (L I Mergaliyeva et al., 2025)(Wang R et al., 2025)(Sharma T et al., 2025)(Febriyanto A et al., 2025)(Shorouk E El-deep et al., 2025)(Al-Rimawi TH et al., 2025)(Bayya AK, 2025)(Hasanudin H, 2025)(Yang Y et al., 2025)(Chemaya N et al., 2025)(Aadithya R Shaji, 2025).

| Percentage of U.S. Adults Using Cryptocurrency | Source |
| --- | --- |
| 7% | Federal Reserve via USAFacts |
| 7% | Federal Reserve via USAFacts |
| 1% | Federal Reserve via USAFacts |
| 1% | Federal Reserve via USAFacts |
| 4% | Federal Reserve via USAFacts |
| 8% | Federal Reserve via USAFacts |
| 4% | Federal Reserve via USAFacts |
| 4% | Federal Reserve via USAFacts |
| 4% | Federal Reserve via USAFacts |
| 4% | Federal Reserve via USAFacts |

*Cryptocurrency Usage Statistics in the United States (2023)*

# VII. Discussion

Individual cryptocurrency use is complex, reaching beyond simple transactions to include social factors, education, and media influence. Various financial platforms and influencers have significantly shaped public opinion and excitement about these currencies. As digital currencies become an alternative asset, media coverage either breeds skepticism or boosts excitement, greatly affecting investment choices. For example, many sensationalized news stories and endorsements from figures on Twitter and YouTube have caused dramatic market swings, drawing in new investors while scaring away others (Onatuyeh E et al., 2025)(T Manoj et al., 2025). Cryptocurrency enthusiasts and self-proclaimed experts often lead social media campaigns that stress potential profits, possibly creating a bandwagon effect encouraging widespread investment without understanding the risks (Joshipura M et al., 2025)(Kou G et al., 2025). The effect of media narratives is significant, acting as vehicles for information and catalysts for emotions, which can increase volatility and speculation among investors (Chaisiripaibool S et al., 2025).

The need for education in the cryptocurrency world is clear. Newcomers often have wrong ideas about the technology behind cryptocurrencies, particularly blockchain and its security and decentralization. Educational efforts, such as those from online platforms, are key to clarifying these concepts and giving people the knowledge to navigate the market. Courses like Learn Cryptocurrency and Crypto Trading for Beginners (Yang J et al., 2025) and Cryptocurrency Trading Education for Beginners (Adamyk B et al., 2025) provide structured learning that covers many aspects of digital currencies. Also, resources like the "Learning Crypto and Bitcoin" course (L I Mergaliyeva et al., 2025) help create a more informed investor base that can make sound decisions based on understanding, not hype. Without this knowledge, people may fall victim to market manipulation and lack the ability to assess investment opportunities properly.

Also, the role of influencers in cryptocurrency spaces often relates to financial literacy. Some influencers promote responsible investing, but others might unintentionally lead their audiences to make impulsive financial choices based on FOMO or excessive optimism. Because of this, viewers should critically assess the information from these figures, balancing excitement with a healthy dose of skepticism (Wang R et al., 2025)(Sharma T et al., 2025). A growing body of research shows that education is crucial for improving personal financial literacy and for creating a healthy crypto market that values informed decision-making (Febriyanto A et al., 2025)(Shorouk E El-deep et al., 2025). With misinformation spreading quickly, people should try to find reliable sources.

Overall, in the cryptocurrency world, passively consuming information can have bad results. Therefore, actively engaging with educational content and critically examining media portrayals are essential for responsible participation. As potential investors learn from structured courses and reliable sources (Al-Rimawi TH et al., 2025)(Bayya AK, 2025), they reduce their risk of speculative bubbles and improve their chances for long-term gains. When this education combines with careful analysis of media, it allows individuals to participate in the cryptocurrency market thoughtfully and strategically. As the market grows, ongoing discussions about improving financial literacy will be important, addressing the mechanics of cryptocurrencies and their broader societal effects (Hasanudin H, 2025)(Yang Y et al., 2025). So, as cryptocurrencies become more popular, promoting education and responsible media consumption will be key to shaping informed individual participation in this innovative financial landscape (Chemaya N et al., 2025)(Aadithya R Shaji, 2025)(I I Supianti et al., 2025)(Jayric A Cinco et al., 2025). This discussion will lead to greater understanding and responsible cryptocurrency usage, increasing their acceptance in mainstream finance.

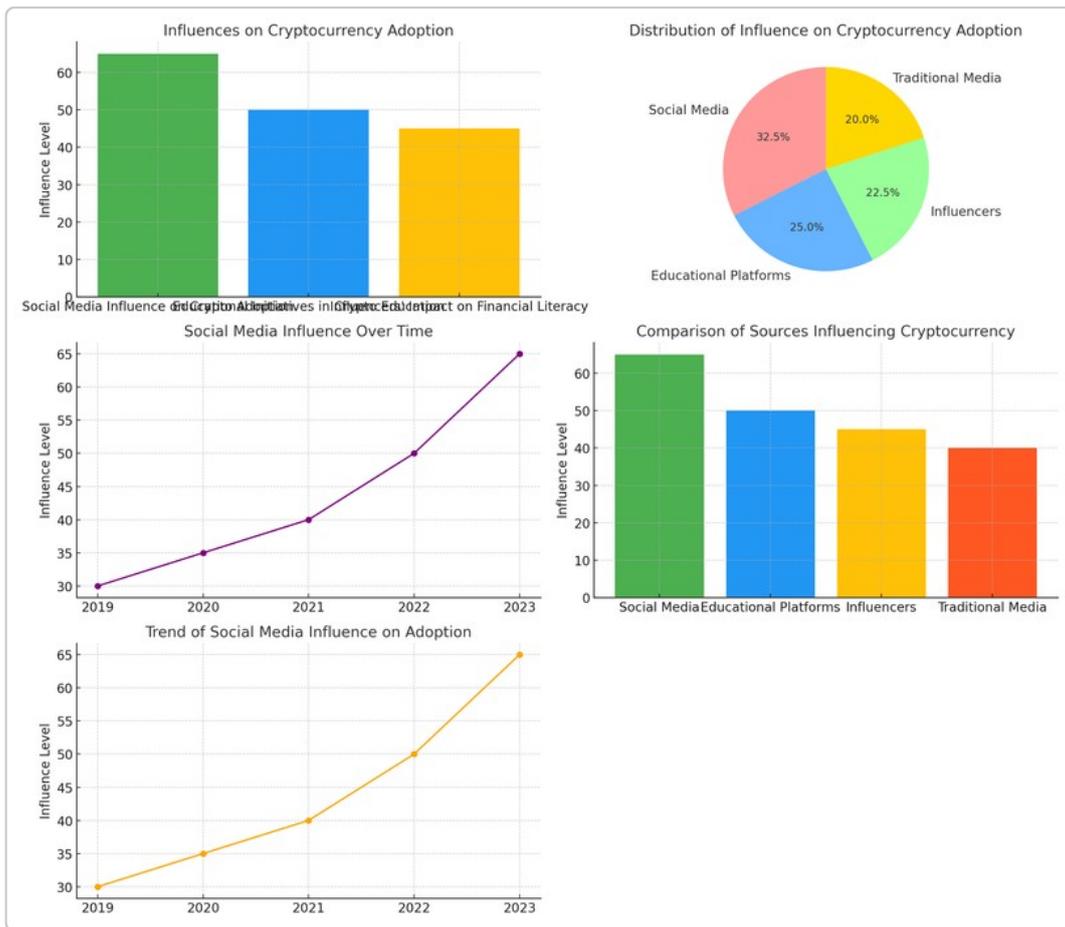

*The charts illustrate the varying influences on cryptocurrency adoption.*

*1. The bar charts depict how social media, educational initiatives, and influencers impact adoption, with social media showing the strongest influence.*
*2. The pie chart reveals that social media and educational platforms dominate the influence landscape.*
*3. The line graphs indicate a consistent increase in the influence of social media over the past five years, highlighting a growing trend in its role in cryptocurrency adoption.*

# VIII. Conclusion

A deep dive into how individuals are using cryptocurrency today reveals a somewhat tangled web of different driving forces all coming together. As previously mentioned, the media and prominent figures have a big impact on how the public views cryptocurrency and how they invest in it. This influence has created a situation where cryptocurrency is seen not just as a tech advancement, but also as a social trend fueled by stories spread through various media outlets. Often, people are drawn in by promises of huge profits and the excitement of being part of a financial revolution. Indeed, smart marketing and endorsements from well-known people have definitely boosted interest in cryptocurrencies, as shown by several studies that point to a link between mainstream media coverage and jumps in cryptocurrency prices (Onatuyeh E et al., 2025)(T Manoj et al., 2025)(Joshipura M et al., 2025). Because of this, individual users often find themselves trying to make their way through a world full of expectations that might not always match up with reality.

Furthermore, education is incredibly important in this constantly changing field. As more and more people get interested in cryptocurrencies, there's a growing need for clear, easy-to-understand educational resources that explain the ins and outs of blockchain tech and cryptocurrency trading. Good education is crucial because it helps people make smart choices and lowers the risks that come with the ups and downs of the cryptocurrency markets. Lots of platforms have popped up to meet this need, offering courses that cover everything from the basics to advanced trading strategies (Kou G et al., 2025)(Chaisiripaibool S et al., 2025)(Yang J et al., 2025). For example, Quttainah (2025) offers thorough intro courses for beginners, making sure that newcomers can grasp the basic mechanics of cryptocurrency (Adamyk B et al., 2025). Similarly, platforms like Cryptocourse.ac provide structured learning paths to help people understand how the market works, which is key to avoiding uninformed investment risks (L I Mergaliyeva et al., 2025). These educational tools not only encourage better investment habits but also promote responsible participation in the cryptocurrency world.

Despite the growing excitement around cryptocurrency, it's important to proceed with caution. The unstable and speculative nature of these digital assets can lead to significant financial risks, particularly for those who aren't prepared. A lack of thorough understanding and blindly trusting media stories may lead individuals to make bad investments. Studies show that many potential investors actually have very little knowledge of blockchain tech and how cryptocurrencies work (Wang R et al., 2025)(Sharma T et al., 2025). The media and influencers can make this knowledge gap even worse, as sensationalized stories might distort how stable and promising cryptocurrencies seem, overshadowing the need for careful consideration and informed decision-making (Febriyanto A et al., 2025)(Shorouk E El-deep et al., 2025). Therefore, encouraging a culture of education and critical thinking is essential for individuals to successfully navigate the complex world of cryptocurrency.

In conclusion, how individuals are currently using cryptocurrencies reflects a mix of enthusiasm, skepticism, and a real need for education. As more people want to get involved in this digital asset world, it's increasingly important to both educate them and promote responsible investment practices. Influencers and media personalities will likely continue to shape opinions on cryptocurrencies; so, combining educational efforts with these influences could lead to a more informed investing public. It's through this educational effort that we can truly unlock the potential of cryptocurrencies, ensuring that users are not only equipped to take advantage of opportunities but also protected from the inherent risks of a quickly evolving financial landscape (Al-Rimawi TH et al., 2025)(Bayya AK, 2025)(Hasanudin H, 2025)(Yang Y et al., 2025) (Chemaya N et al., 2025)(Aadithya R Shaji, 2025)(I I Supianti et al., 2025)(Jayric A Cinco et al., 2025). How we address these issues together will ultimately determine the future of cryptocurrency adoption and usage in the years ahead.

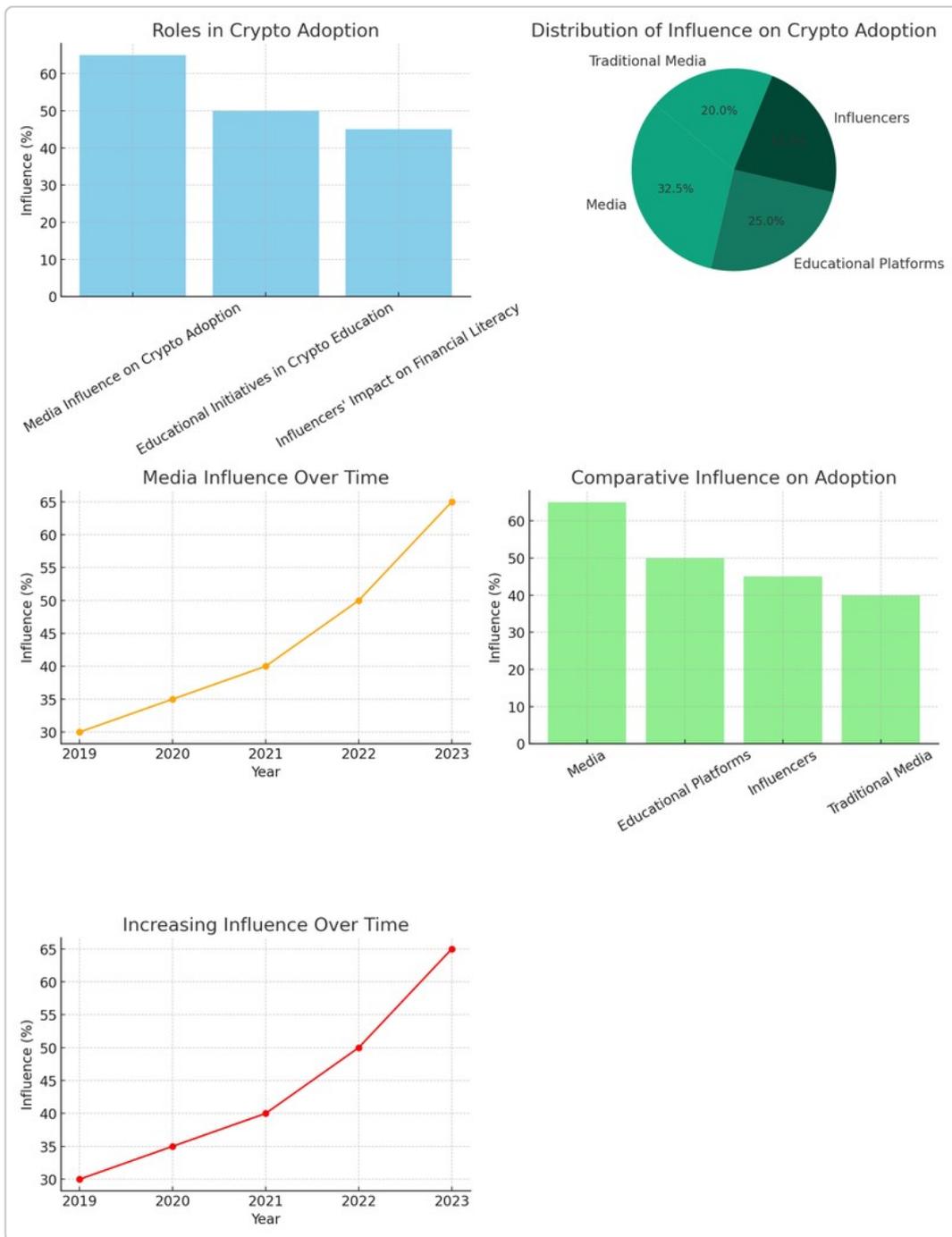

*The charts display various insights into the influence of media, educational initiatives, and influencers on cryptocurrency adoption and financial literacy.*

1. **Bar Chart (Roles in Crypto Adoption):** Highlights the significant contributions of media and educational initiatives.
2. **Pie Chart (Distribution of Influence):** Illustrates the influence distribution among different sources, with media and educational platforms being the most prominent.
3. **Line Chart (Media Influence Over Time):** Shows the upward trend of media influence on cryptocurrency adoption from 2019 to 2023.
4. **Bar Chart (Comparative Influence on Adoption):** Compares the influence of media, educational platforms, influencers, and traditional media.
5. **Line Chart (Increasing Influence Over Time):** Further emphasizes the trend of growing media influence over the same five years.